\documentclass[reprint,onecolumn,aip,jmp,11pt]{revtex4-1}
\usepackage{amsmath}
\usepackage{amsfonts}
\usepackage{amssymb}
\usepackage{graphicx}
\usepackage{dcolumn}
\usepackage{bm}
\usepackage[colorlinks = true, allcolors = blue]{hyperref}
\usepackage{float}
\newcommand{\bd}{\begin{document}}
\newcommand{\ed}{\end{document}}
\newcommand{\bc}{\begin{center}}
\newcommand{\ec}{\end{center}}
\newcommand{\bfr}{\begin{flushright}}
\newcommand{\efr}{\end{flushright}}
\newcommand{\lt}{\left}
\newcommand{\rt}{\right}
\newcommand{\vs}{\vspace}
\newcommand{\hs}{\hspace}
\newcommand{\beq}{\begin{equation}}
\newcommand{\eeq}{\end{equation}}
\newcommand{\lb}{\linebreak}
\newcommand{\pb}{\pagebreak}
\newcommand{\mb}{\makebox}
\newcommand{\fb}{\framebox}
\newcommand{\mc}{\multicolumn}
\newcommand{\ben}{\begin{enumerate}}
\newcommand{\een}{\end{enumerate}}
\newcommand{\bit}{\begin{itemize}}
\newcommand{\eit}{\end{itemize}}
\newcommand{\oln}{\overline}
\newcommand{\un}{\underline}
\newcommand{\lefq}{\lefteqn}
\newcommand{\ba}{\begin{array}}
\newcommand{\ea}{\end{array}}
\newcommand{\beqa}{\begin{eqnarray}}
\newcommand{\eeqa}{\end{eqnarray}}
\newcommand{\beqas}{\begin{eqnarray*}}
\newcommand{\eeqas}{\end{eqnarray*}}
\newcommand{\bfg}{\begin{figure}}
\newcommand{\efg}{\end{figure}}
\newcommand{\bds}{\begin{displaymath}}
\newcommand{\eds}{\end{displaymath}}
\newcommand{\btb}{\begin{tabbing}}
\newcommand{\etb}{\end{tabbing}}
\newcommand{\para}{\parallel}
\newcommand{\pad}{\partial}
\newcommand{\nn}{\nonumber}
\newcommand{\la}{\leftarrow}
\newcommand{\ra}{\rightarrow}
\newcommand{\lgla}{\longleftarrow}
\newcommand{\lgra}{\longrightarrow}
\newcommand{\La}{\Leftarrow}\newcommand{\Ra}{\Rightarrow}
\newcommand{\Lra}{\Leftrightarrow}
\newcommand{\Lgla}{\Longleftarrow}
\newcommand{\Lgra}{\Longrightarrow}
\newcommand{\lan}{\langle}
\newcommand{\ran}{\rangle}
\renewcommand{\a}{\alpha}
\renewcommand{\b}{\beta}
\newcommand{\g}{\gamma}
\newcommand{\G}{\Gamma}
\renewcommand{\d}{\delta}
\newcommand{\eps}{\epsilon}
\newcommand{\Th}{\Theta}
\newcommand{\s}{\sigma}
\newcommand{\lam}{\lambda}
\newcommand{\D}{\Delta}
\newcommand{\ds}{\displaystyle}
\newcommand{\vare}{E}
\newcommand{\pr}{\prime}
\newcommand{\ro}{\rho}
\newcommand{\nab}{\nabla}
\newcommand{\m}{\mu}
\newcommand{\n}{\nu}
\newcommand{\Sg}{\Sigma}
\newcommand{\p}{\pi}
\newcommand{\R}{I\!\!R}
\newcommand{\om}{\omega}
\newcommand{\Om}{\Omega}
\newcommand{\ovra}{\overrightarrow}
\newcommand{\ze}{\zeta}
\newcommand{\vart}{\vartheta}
\newcommand{\tri}{\triangle}
\newcommand{\f}{\frac}
\newcommand{\iny}{\infty}
\newcommand{\pro}{\propto}
\renewcommand{\arraystretch}{1.25}
\begin{document}
\title{Generalized harmonic confinement of massless Dirac fermions in $(2+1)$ dimensions}
\author{Dai-Nam Le}
 \email{ledainam@tdt.edu.vn}   
 \affiliation{Atomic~Molecular~and~Optical~Physics~Research~Group, Advanced Institute of Materials Science, Ton Duc Thang University, Ho~Chi~Minh~City, Vietnam}
 \affiliation{Faculty of Applied Sciences, Ton Duc Thang University, Ho~Chi~Minh~City,~Vietnam}
\author{Van-Hoang Le}
\email{hoanglv@hcmue.edu.vn}
 \affiliation{Department of Physics, Ho Chi Minh City University of Education,\\ 280~An Duong Vuong St.,~Dist. 5,~Ho Chi Minh City,~Vietnam}
\author{Pinaki Roy}
\email{pinaki@isical.ac.in}
\affiliation{Physics and Applied Mathematics Unit, Indian Statistical Institute, Kolkata-700108, India}
\affiliation{Atomic~Molecular~and~Optical~Physics~Research~Group,~Ton~Duc~Thang~University,~Ho~Chi~Minh~City,~Vietnam}
\begin{abstract}
\begin{center}
{\bf Abstract}
\end{center}
In this article we discuss generalized harmonic confinement of massless Dirac fermions in $(2+1)$ dimensions using smooth finite magnetic fields. It is shown that these types of magnetic fields lead to conditional confinement, that is confinement is possible {\it only} when the angular momentum (and parameters which depend on it) assumes some specific values. The solutions for non zero energy states as well as zero energy states have been found exactly.
\end{abstract}
\date{\today}

\maketitle
\section{Introduction}In recent years the massless Dirac equation in $(2+1)$ dimensions has drawn a lot of interest. One of the main reason for this interest is that the dynamics of quasiparticles in materials like graphene is governed by $(2+1)$ dimensional massless Dirac equation except that the quasiparticles move with Fermi velocity $v_F (=10^6~m/s)$ instead of the velocity of light $c$ \cite{g1,g2}. A major challenge in this field is about confining the quasiparticles. The quasiparticles in graphene or similar material can be confined using different methods e.g, by modulated Fermi velocity \cite{fv1,fv2,fv3}, electrostatic fields \cite{el1,zero1,zero3,zero4,zero5}, magnetic fields etc. In particular, magnetic confinement of quasiparticles is possible using, for example, square well magnetic barrier \cite{martino1,martino2}, radial magnetic field \cite{maksym}, decaying gaussian magnetic field \cite{tkg}, hyperbolic magnetic fields \cite{murguia,qesgraphene}, inhomogeneous radial magnetic fields \cite{masir,pratim,downing1,downing2}, one dimensional magnetic fields producing solvable systems  \cite{kuru,midya}, polar angle dependent magnetic fields \cite{jakub}, circular step magnetic field profile \cite{peeters} etc. Among the different types of magnetic fields mentioned above, there are some smooth inhomogeneous magnetic fields \cite{pratim,downing1,downing2} for which the pseudospinor components satisfy Schr\"odinger like equations with quasi exactly solvable effective potentials \cite{tur}. For such magnetic fields the quasiparticles are confined for some values of the angular momentum while for other values the quasiparticles are not bound. In this context it may be noted that there are many ways to produce inhomogeneous magnetic field profiles e.g, using ferromagnetic materials \cite{inhomo1}, non planar substrate \cite{inhomo2}, integrating superconducting elements \cite{inhomo3} etc. 

In this paper our objective is to search for smooth finite magnetic fields which would produce generalized harmonic confinement of quasiparticles i.e, inhomogeneous magnetic fields producing confinement with oscillator spectrum. More precisely, for such magnetic fields one of the pseudospinor components satisfies a Schr\"odinger like equation with the effective potential of a two dimensional oscillator while the other component satisfies a similar equation with the effective potential of a non polynomial oscillator or more general potentials expressed in terms of confluent hypergeometric functions. It will be shown that such confinement is possible only when the parameters of the model assume some particular value(s). In other words, the effective potentials are conditionally exactly solvable \cite{dutra,junker}. The organization of the paper is as follows: in section \ref{forma} we present the model; in section \ref{CES} we shall present several magnetic fields which produce conditional confinement of quasiparticles and finally section \ref{con} is devoted to a conclusion.
 \section{Formalism}\label{forma}
The dynamics of the quasiparticles in graphene is governed by the Hamiltonian
\begin{eqnarray}
\label{eqn:Hamilton}
H = v_F \vec{\sigma} \cdot \vec{\pi} && = v_F \vec{\sigma} \cdot \left( \vec{p} + \vec{A} \right) = v_F \left( 
\begin{matrix}
0 & \Pi _{-} \\
\Pi _+ & 0
\end{matrix} \right) ,
\end{eqnarray}
where $v_F$ is the Fermi velocity, $\sigma = ( \sigma _x , \sigma _y )$ are Pauli matrices, and
\begin{equation}
\Pi_{\pm}= \pi _x \pm i \pi _y = (p_x + A_x) \pm i (p_y + A_y).
\end{equation}
We would like to mention that all equations in the paper are in dimensionless form. Explicitly, the units of length, energy and magnetic field strength we use are respectively $a = 2.46 \; \AA$ (graphene's lattice constant), $E_0 = \hbar c / a = 802 \text{ eV}$ and $B_0 = \hbar c / e a^2 = 1.09 \times 10 ^4 \text{ T}$ which is equivalent to setting $\hbar = c = e = a = 1$.

We choose vector potentials to be of the form
\begin{equation}\label{vector}
A_x=yf(r),~~~~A_y=-xf(r),
\end{equation}
where the form of the function $f(r)$ will be specified later. The magnetic field corresponding to the vector potentials (\ref{vector}) is given by
\begin{equation}\label{defmag}
B_z=-2f(r)-rf'(r).
\end{equation} 
The eigenvalue equation corresponding to the Hamiltonian (\ref{eqn:Hamilton}) is 
\beq
\left(\ba{cc} 0 & \Pi_-\\ \Pi_+ & 0 \ea\right) \psi = E \psi,  
\eeq
where $\psi=(\psi_1,\psi_2)^T$ is a two component pseudospinor. The above equation can be written in terms of the components as
\beq\label{coupled}
\ba{l}
\Pi _{-} \psi_2 = \epsilon \psi_1 ,\\
\Pi _{+} \psi_1 = \epsilon \psi_2 ,
\ea
\eeq 
where $\epsilon=E/v_F$. Since the magnetic field is a radial one, the pseudospinor components can be taken as
\begin{equation}\label{radial}
\psi_1=e^{im\theta}r^{-1/2}\phi_1(r),~~~~\psi_2=e^{i(m+1)\theta}r^{-1/2}\phi_2(r),
\end{equation}
where $m=0,\pm 1,\pm 2,\cdots$ denotes the angular momentum quantum number. Now using (\ref{radial}) the intertwining relations (\ref{coupled}) can be written in terms of polar coordinates as
\begin{equation}\label{intert}
\begin{array}{l}
\displaystyle\left( \dfrac{\partial}{\partial r}- \dfrac{m+\frac{1}{2}}{r} + r f \right) \phi _1 = i \epsilon \phi _2,\\
\displaystyle\left( \dfrac{\partial}{\partial r}+ \dfrac{m+\frac{1}{2}}{r} - r f \right) \phi _2 = i \epsilon \phi _1.
\end{array}
\end{equation}
Now eliminating $\phi_1$ in favor of $\phi_2$ (and vice-versa), the equations for the components can be written as
\begin{equation}\label{eqn:psi1-a}
\displaystyle\left[-\frac{d^2}{dr^2}+\frac{m^2-\frac{1}{4}}{r^2}+ r^2 f^2 - 2(m+1)f - rf'\right] \phi_1 = \epsilon^2 \phi_1,\\
\end{equation}
\begin{equation}\label{eqn:psi2-a}
\displaystyle\left[-\frac{d^2}{dr^2} + \frac{(m+1)^2-\frac{1}{4}}{r^2}+ r^2 f^2 - 2mf + rf'\right]   \phi _2 = \epsilon^2 \phi _2. 
\end{equation}
In the next section we shall find magnetic fields for which at least one of the above equations becomes conditionally exactly solvable.

\section{Generalized harmonic confinement}\label{CES} 
Here we shall consider several magnetic field profiles for which all or some bound state solutions can be found {\it only} when the parameters of the model assume some specific values. To this end we choose the function $f(r)$ to be of the form
\begin{equation}
\label{f}
f(r) = \dfrac{\lambda}{2} + g(r),
\end{equation}
where $\lam$ is a constant and the function $g(r)$ will be determined later. Then the resulting magnetic field is given by
\begin{equation}
\label{eqn:consider-B}
B_z (r) = - \lambda - 2g(r)-rg'(r).
\end{equation}
Note that when $g(r)=0$, the magnetic field becomes a homogeneous one and the effective potentials in Eqs.(\ref{eqn:psi1-a}) and (\ref{eqn:psi2-a}) are just those corresponding to a two dimensional oscillator. Now using (\ref{f}) the effective potentials in Eqs.(\ref{eqn:psi1-a}) and (\ref{eqn:psi2-a}) can be found to be
\begin{eqnarray}
V_1(r) &&=  \displaystyle\f{m^2-\f{1}{4}}{r^2}+\dfrac{r^2 \lambda ^2}{4} - (m+1) \lambda + \left[ r^2 g^2 + \lambda r^2 g - 2(m+1) g - r g' \right] \label{eqn:V1-g},\\
V_2 (r) &&= \displaystyle\f{(m+1)^2-\f{1}{4}}{r^2}+ \dfrac{r^2 \lambda ^2}{4} - m \lambda + \left[ r^2 g^2 + \lambda r^2 g - 2 m g + r g' \right] .\label{eqn:V2-g} 
\end{eqnarray}
We shall now identify $V_2(r)$ with the effective potential of a two dimensional harmonic oscillator (displaced in the energy scale) and this requires that the function $g(r)$ satisfies the following Riccati equation :
\beq\label{riccati}
r^2 g^2 + \lambda r^2 g - 2 m g + r g' = \mu ,
\eeq
where $\mu$ is a constant. We now linearize Eq.(\ref{riccati}) by putting $\displaystyle g=\f{u'}{ru}$ and obtain
\beq\label{u}
r u'' + \left( \lambda r^2 - 2 m -1 \right) u' - \mu r u = 0.
\eeq
The solution of Eq.(\ref{u}) is given by
\beq\label{solu}
u(r) = c_1 \; {}_1 F_1 \left( - \dfrac{\mu}{2 \lambda} ; - m ; - \dfrac{\lambda r^2}{2} \right)  + c_2 r^{2 (m+1)}\; {}_1 F_1 \left( 1 + m  - \dfrac{\mu}{2 \lambda}; m + 2 ; - \dfrac{\lambda r^2}{2} \right),
\eeq
where $c_{1,2}$ are arbitrary constants and $_1F_1(a,b;x)$ denotes the confluent hypergeometric function \cite{abra}. Having found $u$, it is now easy to obtain the function $\displaystyle g(r)=\f{u'}{ru}$ and in turn the magnetic field. As mentioned before that for this choice of $u(r)$ the potential $V_2(r)$ represents a harmonic oscillator for which the energy and the eigenfunctions are well known. Then using the intertwining relation Eq.(\ref{intert}) one may obtain the component $\phi_1(r)$and hence the complete pseudospinor. Next, we shall consider several possibilities of constructing the magnetic field.

\subsection{\label{sec:2A}$c_1\neq 0, c_2=0$:}

In this case we find
\beq\label{g}
g(r)= \dfrac{\mu \; {}_1 F_1 \left( 1 - \dfrac{\mu}{2 \lambda} ; M+2 ; - \dfrac{\lambda r^2}{2} \right) }{2 (M+1) \; {}_1 F_1 \left( - \dfrac{\mu}{2 \lambda} ; M+1 ; - \dfrac{\lambda r^2}{2} \right)},
\eeq
and the magnetic field is given by
\begin{eqnarray}
\label{eqn:B-1}
B_z (r) =- \lambda \; && \displaystyle - \dfrac{\mu}{M+1} \dfrac{{}_1 F_1 \left( 1 - \f{\mu}{2 \lambda}; M+2 ; - \f{\lambda r^2}{ 2 }\right) }{ {}_1 F_1 \left( - \f{\mu}{2 \lambda}; M+1 ; - \f{\lambda r^2}{ 2} \right)} \nonumber \\
&& + \displaystyle\left[ \dfrac{\mu r}{2 (M+1)}  \dfrac{{}_1 F_1 \left( 1 - \f{\mu }{2 \lambda} ; M+2; - \f{\lambda r^2}{ 2} \right) }{ {}_1 F_1 \left( - \f{\mu}{2\lambda} ; M+1 ; - \f{\lambda r^2}{2}\right)} \right] ^2  \nonumber\\
&&  - \displaystyle \dfrac{\mu (\mu - 2 \lambda ) r^2 }{4 (M+1)(M+2)} \dfrac{{}_1 F_1 \left( 2 -\f{ \mu }{2 \lambda }; M+3 ; - \f{\lambda r^2 }{2} \right) }{ {}_1 F_1 \left( -\f{ \mu }{2 \lambda }; M+1 ; - \f{\lambda r^2 }{2} \right)},
\end{eqnarray}
where $m=-(M+1)$. We would like to point out that in order to produce a non singular magnetic field and effective potential $V_{1,eff}$ the function $g(r)$ should not have any poles. Furthermore, a singular $V_{1,eff}$ will not be isospectral with $V_{2,eff}(r)$. Therefore, it is necessary to choose various parameters in such a way that ${}_1 F_1 \left( -\f{ \mu }{2 \lambda }; M+1 ; - \f{\lambda r^2 }{2} \right)$ does not have a zero.

In this context it may be noted that depending on the parameters the confluent hypergeometric series can either be a finite series or an infinite one. We shall now consider both these possibilities and construct the corresponding magnetic fields by suitably choosing the parameters $\lam, \mu$ and $M$.

\subsubsection{\mbox{\boldmath$M\geq 0,~~\lambda > 0~~\mu = 2 N \lambda,~~ N = 1, 2, \dots $}}
\begin{figure}[H]
\begin{center}
\includegraphics[width = 0.6 \textwidth]{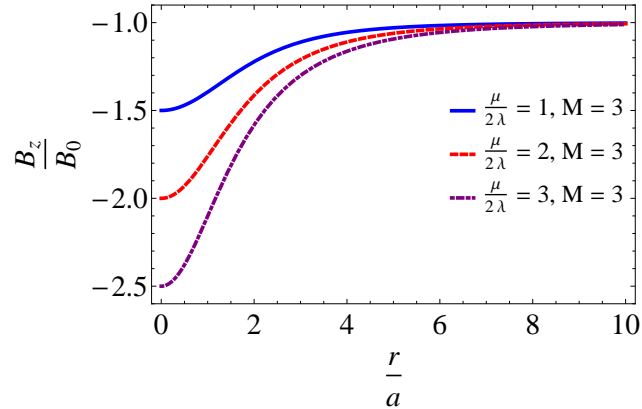}
\caption{\label{fig:B-1}Typical behavior of the magnetic field when $\mu / 2 \lambda = N$ is a positive integer. }
\end{center}
\end{figure}

\begin{figure}[H]
\begin{center}
\includegraphics[width = 0.6 \textwidth]{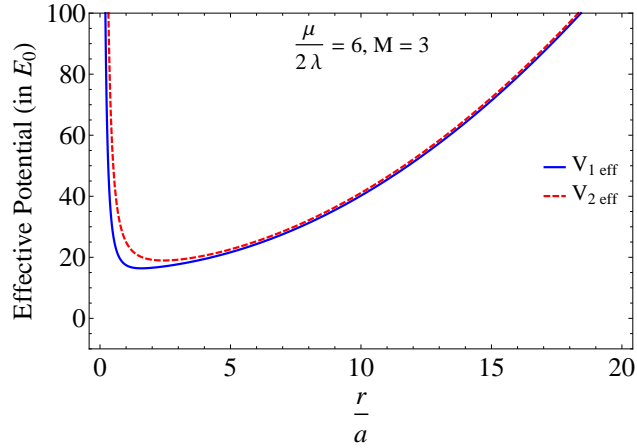}
\caption{\label{fig:V-1}Plot of the effective potentials  $V_{1,eff}$, $V_{2,eff}$ when $\mu / 2 \lambda = N$ is a positive integer. }
\end{center}
\end{figure}

\begin{figure}[H]
\begin{center}
\includegraphics[width = 0.6 \textwidth]{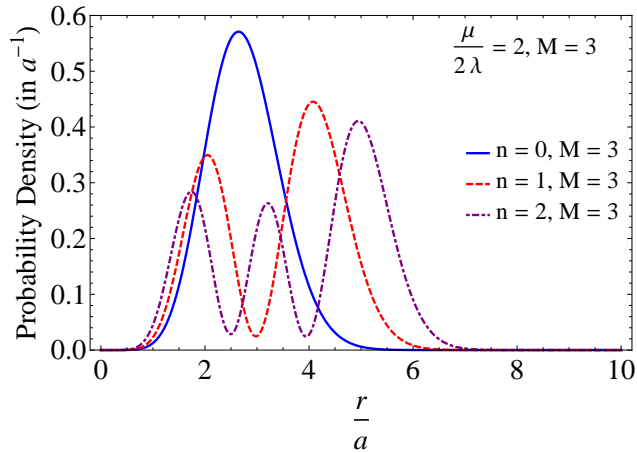}
\caption{\label{fig:rho-1}Plot of the probability density for different quantum numbers.}
\end{center}
\end{figure}

In this case the confluent hypergeometric series ${}_1F_1(-N,M+1,-\lam r^2/2)$ becomes a finite series and can be expressed in terms of the associated Laguerre polynomials $L ^{\alpha}_n (x)$ . Consequently $g(r)$ can be written as
\begin{eqnarray}
\label{eqn:g-1-N}
g(r) = \dfrac{\lambda L^{M+1}_{N-1} ( - \lambda r^2 / 2)}{L^{M}_{N} ( - \lambda r^2 / 2)}.
\end{eqnarray}
Note that if we denote the roots of $L_N^M(x)$ by $x_i$, then $g(r)$ can be written as
\beq
g(r)=\sum_{i=0}^N\frac{2g_i}{1+g_ir^2},
\eeq
where $x_i=\lam/2g_i,~~g_i>0$. Then from Eq.(\ref{defmag}) the corresponding magnetic field $B_z (r)$ can be found to be:
\begin{equation}
\label{eqn:B-1-N}
B_z(r)=-\lam-\sum_{i=0}^N\f{2g_i}{1+g_ir^2}.
\end{equation}
We note that magnetic fields of the type (\ref{eqn:B-1}) was considered previously for some particular values of $N$ in Ref. \onlinecite{roy17} and to obtain results for any particular value of $N$, one just has to put that value of $N$ in our general results. From (\ref{eqn:B-1-N}) it is observed that the magnetic field is non singular with a maximum value $-\lam$ and a minimum of $-\lam-4\sum_{i=0}^N g_i$. A plot of the magnetic fields for different parameter values is shown in Fig \ref{fig:B-1}.
Before we proceed to obtain the spectrum, let us note that because of the choice of the magnetic field the effective potential $V_{1,eff}$ in (\ref{eqn:V2-g}) is always that of a two dimensional harmonic oscillator one (displaced in the energy scale) but the effective potential $V_{2,eff}$ in (\ref{eqn:V1-g}) is in general a non polynomial oscillator type although they are isospectral. In Fig \ref{fig:V-1} we have presented plots of these potentials for some admissible parameter values.
We shall now obtain the spectrum. To this end we note that Eq.\eqref{eqn:psi2-a} for the lower component $\phi_2$ becomes:
\begin{equation}
\label{eqn:general-2-harmo-b}
\left[ - \dfrac{d^2}{dr^2} + \dfrac{M^2 -1/4}{r^2} + \dfrac{\lambda ^2 r^2}{4} \right] \phi _2 = \left[ \epsilon ^2 - (M + 1 + 2 N) \lambda \right] \phi _2.
\end{equation}
Eq.(\ref{eqn:general-2-harmo-b}) is the eigenvalue equation of a two-dimensional isotropic harmonic oscillator whose solutions are well known. Using these results the eigenvalues $E_{n,M}$ and the corresponding wave functions $\phi _{2,n,M}$ can be found to be:
\begin{equation}
\label{eqn:E-N-harmo}
E_{n,M} = v_F \epsilon _{n,M} =  \pm v_F \sqrt{2 \lambda (n+M+1+N)}, \quad ~n=0,1,2,\dots;M=0,1,2,\cdots ,
\end{equation}
and
\begin{equation}
\label{eqn:phi2-N-harmo}
\phi _{2,n,M} (r) \sim r^{M+1/2} e^{- \lambda r^2 / 4} L _{n}^{M} \left( \lambda r^2 / 2 \right).
\end{equation}
The upper component $\phi_{1,n,M}$ can now be obtained using the intertwining relation \eqref{intert} and is given by
\beq\label{phi1}
\phi_{1,n,M}(r)=i \lam\epsilon _{n,M}^{-1} \f{r^{M+1} e^{- \lambda r^2 / 4}}{L_{N}^{M} ( - \lambda r^2 / 2)} \left[L_{N}^{M} ( - \lambda r^2 / 2) L _{n}^{M+1} \left( \lambda r^2 / 2 \right) + L_{N-1}^{M+1} ( - \lambda r^2 / 2) L_n^M( \lambda r^2 / 2)\right].
\eeq
It is interesting to note that the above pseudospinor component can be written in terms of the type I $X_m$ exceptional Laguerre polynomials $\hat{L}^k_{n,m}(r^2)$. Using the relation \cite{eop1}
\beq
\hat{L}^k_{n,m}(r^2)= L_m^k(-r^2)L^{k-1}_{n-m}(r^2) + L_m^{k-1}(-r^2)L^{k}_{n-m-1}(r^2),  \ \ \  n\ge m
\eeq
the upper component can be written as
\beq
\phi_{1,n,M}(r)=i \lam\epsilon _{n,M}^{-1} \f{r^{M+1} e^{- \lambda r^2 / 4}}{L_{N}^{M} ( - \lambda r^2 / 2)} {\hat L}^{M+1}_{N+n,n}(-\lam r^2/2).
\eeq
Thus the pseudopspinor $\psi _{n,M} (r, \theta )$ is given by:
\begin{eqnarray}
\label{eqn:psi-N-harmo}
\psi _{n,M} (r , \theta) \sim e^{- i (M+1) \theta} \times \f{r^{M}e^{- \lambda r^2 / 4} }{L_{N}^{M} ( - \lambda r^2 / 2)} \times  \left(
\begin{matrix}
i \epsilon _{n,M}^{-1} \lambda r {\hat L}^{M+1}_{N+n,n}(-\lam r^2/2) \\
e^{i \theta}{L_{N}^{M} ( - \lambda r^2 / 2)}L _{n}^{M} \left( \lambda r^2 / 2 \right) 
\end{matrix}
\right).
\end{eqnarray}
A plot of the probability density for different quantum numbers is shown Fig \ref{fig:rho-1}. 
Let us now examine degeneracy of the eigenvalues. From Eq.\eqref{eqn:E-N-harmo}, it is easy to find that the energy values depend explicitly on the sum $n+M$. Thus the ground state $E_{0,0}$ is non-degenerate while the degeneracy of the excited state $(n,M)$ is $n+M+1$. 

\subsubsection{\mbox{\boldmath$M<0,~~\lambda <0,~~\mu = 2 N \lambda,~~ N = 1, 2, \dots $}}
\begin{figure}[H]
\begin{center}
\includegraphics[width = 0.6 \textwidth]{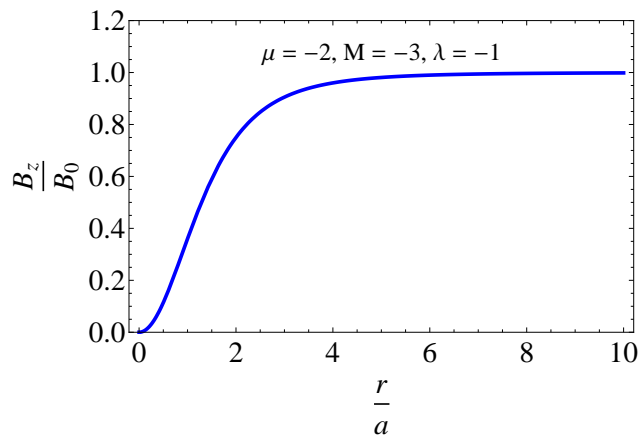}
\caption{\label{fig:B-2}Shape of the magnetic field when $\lam = -1, \mu = -2, M = - 3$.}
\end{center}
\end{figure}

\begin{figure}[H]
\begin{center}
\includegraphics[width = 0.6 \textwidth]{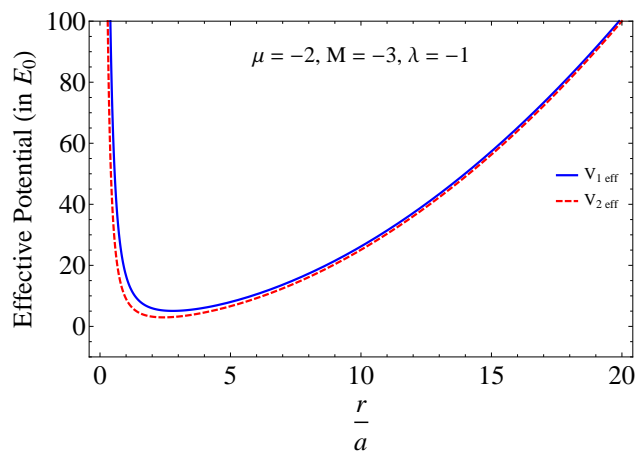}
\caption{\label{fig:V-2}Effective potentials $V_{1,eff}$ and $V_{2,eff}$ when $\lam = -1, \mu = -2, M = - 3$.}
\end{center}
\end{figure}

In the last section it was shown that confinement is possible for $M=0,1,2,\cdots$ i.e, $m=-1,-2,\cdots$. Here it will be shown that confinement is possible even when the angular momentum $m$ is positive. To do this we note that the confluent hypergeometric series ${}_1F_1(a,b;x)$ has poles when $b$ is a negative integer. However, it is interesting to note that the series actually terminates even when both $a$ and $b$ are negative integers and $a>b$. Rather than going into the mathematical details of this interesting topic \cite{negative,sansone}, we just show that with judicious choice of the parameters it is possible to construct finite magnetic fields even when the angular momentum $m$ is positive. Thus we consider the parameter values $\lam = -1, \mu = -2, M = - 3 (m=2)$. In this case the solution of Eq.(\ref{u}) is given by
\beq\label{u1}
u(r)=1+\f{r^2}{4} .
\eeq  
The magnetic field and the effective potentials are given respectively by
\begin{eqnarray}
B_z(r)&=&\f{r^2(r^2+8)}{(r^2+4)^2}, \label{mag1}\\
V_{1,eff}(r)&=&\f{r^2}{4}+1+\f{15}{4r^2}+\f{4}{r^2+4}-\f{20}{(r^2+4)^2}, \\
V_{2,eff}(r)&=&\f{r^2}{4}+\f{35}{4r^2} .
\end{eqnarray}
Plots of the magnetic field and the effective potentials are shown in Figs. \ref{fig:B-2} and \ref{fig:V-2}.

The bound state energy values are given by
\begin{equation}
E_{n,-3} =  \pm v_F \sqrt{2n}, \quad ~n=0,1,2,3,\dots ,
\end{equation}
It is seen that the lowest energy value is zero and the corresponding pseudospinor is a singlet of the form
\begin{eqnarray}
\psi _{0,-3} (r , \theta) \sim e^{ 4 i  \theta}r^{3} e^{- r^2 / 4}\left(
\begin{matrix}
0 \\
e^{i \theta} L _{1}^{3} \left( r^2 / 2 \right) 
\end{matrix}
\right).
\end{eqnarray}
On the other hand the pseudospinors corresponding to the non zero energy levels are given by
\begin{eqnarray}\label{eqn:psi-2}
\psi _{n,-3} (r , \theta) \sim e^{ 4 i  \theta}r^{2} e^{- r^2 / 4}\left(
\begin{matrix}
\displaystyle\pm \f {i(r^4-12r^2-12)}{2(r^2+1)\sqrt{n}}L^{4}_{n+1}( r^2/2)  \\
r e^{i \theta} L _{n+1}^{3} \left( r^2 / 2 \right) 
\end{matrix}
\right),~n=1,2,\cdots.
\end{eqnarray}

\begin{figure}[H]
\begin{center}
\includegraphics[width = 0.6 \textwidth]{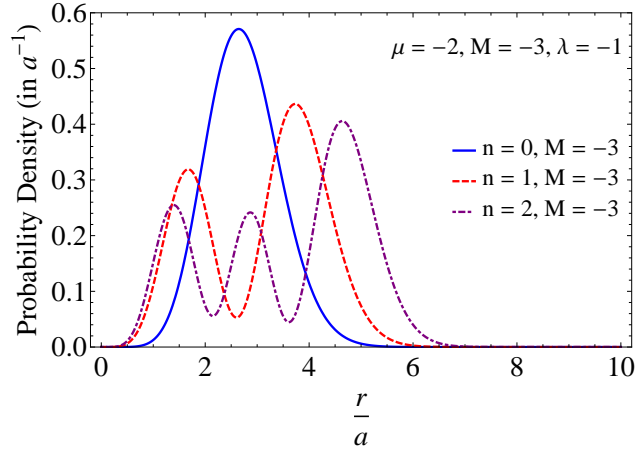}
\caption{\label{fig:rho-2}The probability density when $\lam = -1, \mu = -2, M = - 3$ for $n = 1, 2, 3$ states.}
\end{center}
\end{figure}

A plot of probability density for some of the states is shown in Fig. \ref{fig:rho-2}.

\subsubsection{\mbox{\boldmath$M>0,~~\lambda > 0,~~\mu/2\lam\neq$ positive integer}}
\begin{figure}[H]
\begin{center}
\includegraphics[width = 0.6 \textwidth]{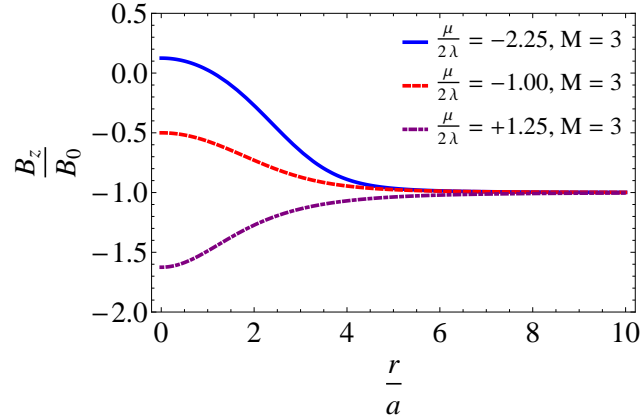}
\caption{\label{fig:B-3}Typical magnetic field profile when $\mu / 2 \lambda $ is a positive fraction ($\mu / 2 \lambda = 1.25$), a negative fraction ($\mu / 2 \lambda = -2.25$) and a negative integer ($\mu / 2 \lambda = -1$).} 
\end{center}
\end{figure}

\begin{figure}[H]
\begin{center}
\includegraphics[width = 0.6 \textwidth]{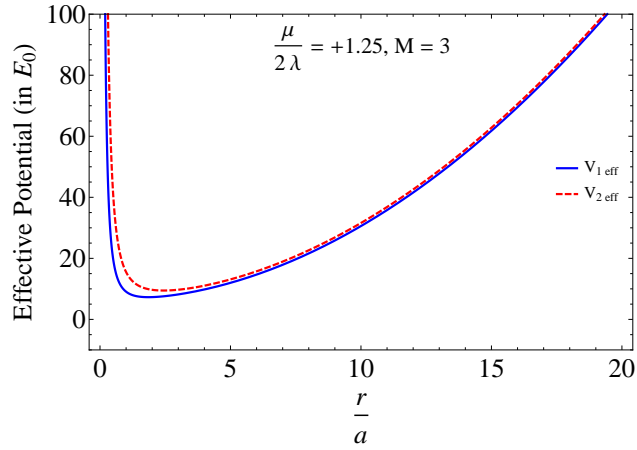}
\caption{\label{fig:V-3}Shape of the effective potentials $V_{1,eff}$ and $V_{2,eff}$ when $\mu / 2 \lambda $ is not a positive integer. }
\end{center}
\end{figure}

This case is the most general one in the sense that confluent hypergeometric series is no longer a finite one and consequently there is more freedom in the choice of the parameters $\lam$ and $\mu$. In this case the function $f(r)$ and the corresponding magnetic field $B_z(r)$ have the same expressions as in (\ref{g}) and (\ref{eqn:B-1}) respectively with the appropriate parameter values as mentioned above. In this section we shall consider two possibilities, namely, $(1)~ \displaystyle\f{\mu}{2\lam}=$ negative integer, $\lam, M>0$ and $(2)~ \displaystyle\f{\mu}{2\lam}=\pm$ fraction, $\displaystyle\lam, M>0$. In Figs. \ref{fig:B-3} and \ref{fig:V-3} we present profiles of the magnetic field and the effective potentials for suitable values of the parameters. For both these cases mentioned above the generic forms of the energy values and the corresponding pseudospinors are given by  
\begin{eqnarray}
\label{eqn:psi-N-c}
&&E_{n,M} =  \pm v_F \sqrt{2 \lambda (n+M+1) + \mu }, \quad ~n=0,1,2,\dots ,\\
&& \psi _{n,M} (r , \theta) \sim r^{M} e^{- \lambda r^2 / 4} e^{- i (M+1) \theta} \times \nonumber\\
&&\quad \times \left(
\begin{matrix}
i \epsilon _{n,M}^{-1} \lambda r \left[ L _{n}^{M+1} \left( \lambda r^2 / 2 \right) + \dfrac{\mu L _{n}^{M} \left( \lambda r^2 / 2 \right)}{2 \lambda (M+1)} \dfrac{{}_1 F_1 \left( 1 - \mu / (2 \lambda ) ; M+2 ; - \lambda r^2 / 2 \right) }{{}_1 F_1 \left( - \mu / ( 2 \lambda ) ; M+1 ; - \lambda r^2 / 2  \right)} \right] \\
L_{n}^{M} \left( \lambda r^2 / 2 \right) e^{i \theta}
\end{matrix}
\right).
\end{eqnarray}

\begin{figure}[H]
\begin{center}
\includegraphics[width = 0.6 \textwidth]{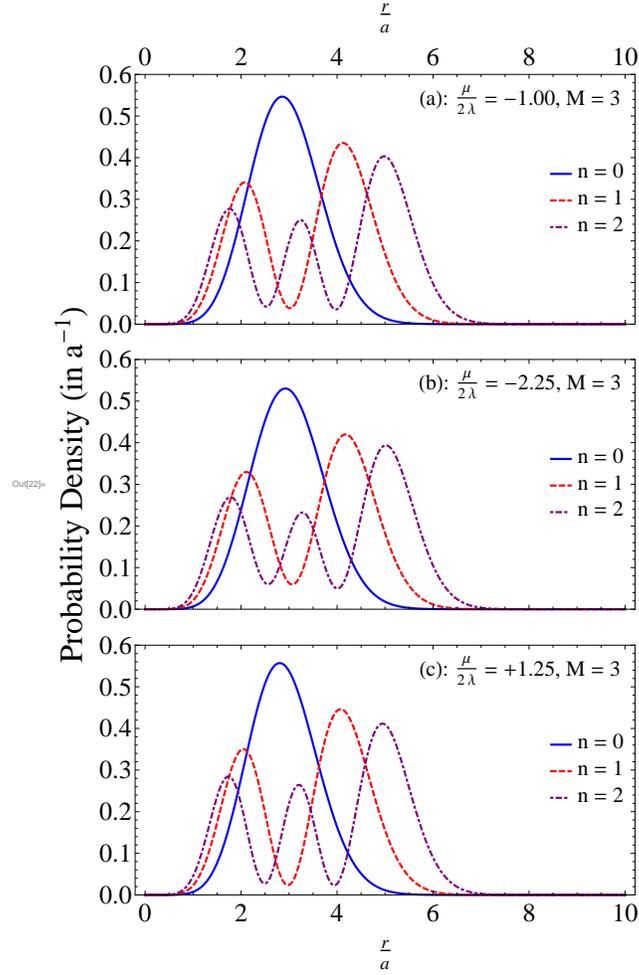}
\caption{\label{fig:rho-3}(a) The probability density when $\mu / 2 \lambda = - 1$ for the states $(n,M) = (0,3)$, $(1,3)$ and $(2,3)$. (b) The probability density when $\mu / 2 \lambda = - 2.25$ for the states $(n,M) = (0,3)$, $(1,3)$ and $(2,3)$. (c) The probability density when $\mu / 2 \lambda = 1.25$ for the states $(n,M) = (0,3)$, $(1,3)$ and $(2,3)$.}
\end{center}
\end{figure}
\subsection{\label{sec:2B}$c_1 = 0, c_2 \neq 0$}
In this case we find
\begin{eqnarray}
\label{eqn:g-2}
g(r) = \dfrac{2 M}{r^2} + \dfrac{ \nu {}_1 F_1 \left( 1 - \nu / 2 \lambda ; M+2 ; - \lambda r^2 / 2 \right)}{2 (M+1) {}_1 F_1 \left( - \nu / 2 \lambda ; M+1 ; - \lambda r^2 / 2 \right) } 
\end{eqnarray}
and the magnetic field is given by
\begin{eqnarray}
\label{eqn:B-2}
B_z (r) = - \lambda \; && \displaystyle - \dfrac{\nu}{M+1} \dfrac{{}_1 F_1 \left( 1 - \f{\nu}{2 \lambda}; M+2 ; - \f{\lambda r^2}{ 2 }\right) }{ {}_1 F_1 \left( - \f{\nu}{2 \lambda}; M+1 ; - \f{\lambda r^2}{ 2} \right)} \nonumber \\
&& + \displaystyle\left[ \dfrac{\nu r}{2 (M+1)}  \dfrac{{}_1 F_1 \left( 1 - \f{\nu }{2 \lambda} ; M+2; - \f{\lambda r^2}{ 2} \right) }{ {}_1 F_1 \left( - \f{\nu}{2\lambda} ; M+1 ; - \f{\lambda r^2}{2}\right)} \right] ^2  \nonumber\\
&&  - \displaystyle \dfrac{\nu (\nu - 2 \lambda ) r^2 }{4 (M+1)(M+2)} \dfrac{{}_1 F_1 \left( 2 -\f{ \nu }{2 \lambda }; M+3 ; - \f{\lambda r^2 }{2} \right) }{ {}_1 F_1 \left( -\f{ \nu }{2 \lambda }; M+1 ; - \f{\lambda r^2 }{2} \right)}  ,
\end{eqnarray}
where the notations $m = M - 1$ ($M = 0, 1, 2, \dots$) and $\nu = \mu - 2 \lambda M$ are used. The magnetic field \eqref{eqn:B-2} in this case is exactly the same as the magnetic field \eqref{eqn:B-1} on section \ref{sec:2A} if $\mu$ were replaced by $\nu$. The reason for this is that the first term $2M/r^2$ of $g(r)$ in \eqref{eqn:g-2} corresponds to a Aharonov-Bohm type vector potential and does not produce any magnetic field strength on the graphene plane.

The conditional harmonic solution for energy $E_{n,M}$ and the lower component $\phi _{2,n,M}$ are given by
\begin{equation}
\label{eqn:E-nu-2}
E_{n,M} = v_F \epsilon _{n,M} = \pm v_F \sqrt{2 \lambda (n+M+1) + \nu}, \quad n = 0, 1, 2, \dots,
\end{equation} 
\begin{equation}
\label{eqn:phi2-nu-2}
\phi _{2,n,M} (r) \sim r^{M+1/2} e^{- \lambda r^2 / 4} \mathcal{L}_n^M ( \lambda r^2 / 2) .
\end{equation}
From the intertwining relation \eqref{intert} the solution $\phi _{1,n,M}$ can be found as:
\begin{equation}
\label{eqn:phi1-nu-2}
\phi _{1,n,M} (r) \sim i \epsilon _{n,M}^{-1} \lambda r^{M+3/2} \left[ L _{n}^{M+1} \left( \lambda r^2 / 2 \right) + \dfrac{\nu L _{n}^{M} \left( \lambda r^2 / 2 \right)}{2 \lambda (M+1)} \dfrac{{}_1 F_1 \left( 1 - \nu / (2 \lambda ) ; M+2 ; - \lambda r^2 / 2 \right) }{{}_1 F_1 \left( - \nu / ( 2 \lambda ) ; M+1 ; - \lambda r^2 / 2  \right)} \right] .
\end{equation}
Hence, the pseudospinor corresponding to the energy (\ref{eqn:phi2-nu-2}) is
\begin{eqnarray}
\label{eqn:psi-nu}
&& \psi _{n,M} (r , \theta) \sim r^{M} e^{- \lambda r^2 / 4} e^{i (M-1) \theta} \times \nonumber\\
&&\quad \times \left(
\begin{matrix}
i \epsilon _{n,M}^{-1} \lambda r \left[ L _{n}^{M+1} \left( \lambda r^2 / 2 \right) + \dfrac{\nu L _{n}^{M} \left( \lambda r^2 / 2 \right)}{2 \lambda (M+1)} \dfrac{{}_1 F_1 \left( 1 - \nu / (2 \lambda ) ; M+2 ; - \lambda r^2 / 2 \right) }{{}_1 F_1 \left( - \nu / ( 2 \lambda ) ; M+1 ; - \lambda r^2 / 2  \right)} \right] \\
L_{n}^{M} \left( \lambda r^2 / 2 \right) e^{i \theta}
\end{matrix}
\right),
\end{eqnarray}
which is only different from case \textbf{A} because of the presence of the factor $e^{i (M-1) \theta}$ factor instead of $e^{-i (M+1) \theta}$. This means the energy and the probability density corresponding the case \textbf{B} are exactly the same as in case \textbf{A}.

\section{Zero energy solutions}
\subsection{$c_1 \neq 0, c_2 = 0$}
In the previous sections we have discussed conditional confinement for non zero energy. On the other hand, zero energy solutions are important too and have been discussed by many authors \cite{zero1,zero3,zero4,zero5}. In this section we shall obtain zero energy solutions of our problem. Such solutions can be obtained from the intertwining relations (\ref{intert}). For $\eps=0$ we obtain from the first relation of (\ref{intert}) and (\ref{solu})
\beq\label{zero1}
\phi_{1,M}(r)\sim r^{-M-1/2}e^{-\lam r^2/4}\f{1}{{}_1F_1\left(-\f{\mu}{2\lam},M+1;-\f{\lam r^2}{2}\right)},~~M = -1,-2,-3\cdots,
\eeq
where $\mu$, $\lambda$ and $M$ are parameters of the magnetic field. The pseudospinor is thus given by
\beq\label{zero11}
E_{0,M}=0,~~~~\psi_{0,M}(r,\theta)\sim  e^{-i(M+1)\theta}\left(\ba{c} \dfrac{r^{-M-1}e^{-\lam r^2/4}}{{}_1F_1\left(-\f{\mu}{2\lam},M+1;-\f{\lam r^2}{2}\right)}\\0\ea\right),~~\lambda >0, ~~ M = -1,-2,-3 \cdots .
\eeq
Similarly another set of zero energy solutions can be obtained from the second equation of (\ref{intert}) and is given by
\beq\label{zero2}
E_{0,M}=0,~~~~\psi_{0,M}(r,\theta)\sim e^{i (M+1)\theta}\left(\ba{c} 0\\r^{M+1} e^{\lam r^2/4}{}_1F_1\left(-\dfrac{\mu}{2 \lambda},M+1; - \f{\lam r^2}{2}\right)\ea\right),~~\lambda < 0,~~M=0,1,2 \cdots ,
\eeq
when $\mu / 2 \lambda \geq - (M+1) $.
Note that in both the cases the energy levels are infinitely degenerate with respect to the quantum number $M$.

\subsection{$c_1 = 0, c_2 \neq 0$}
For $\eps=0$ we obtain from the first relation of (\ref{intert}) and (\ref{solu})
\beq\label{zero3}
\phi_1(r)\sim r^{-M-1/2}e^{-\lam r^2/4}\f{1}{{}_1F_1\left(-\f{\nu}{2\lam},M+1;-\f{\lam r^2}{2}\right)},~~M = -1, -2, -3 \cdots,
\eeq
where $\nu$, $\lambda$ and $M$ are parameters of the magnetic field. The pseudospinor is thus given by
\beq
E_{0,M}=0,~~~~\psi(r,\theta)\sim  e^{i(M-1)\theta}\left(\ba{c} \dfrac{r^{-M-1}e^{-\lam r^2/4}}{{}_1F_1\left(-\f{\nu}{2\lam},M+1;-\f{\lam r^2}{2}\right)}\\0\ea\right),~~\lambda >0, ~~ M = -1, -2, -3 \cdots .
\eeq

Similarly another set of zero energy solutions can be obtained from the second equation of (\ref{intert}) and is given by
\beq\label{zero4}
E_{0,M}=0,~~~~\psi_{0,M}(r,\theta)\sim e^{- i (M-1) \theta}\left(\ba{c} 0\\r^{M+1} e^{\lam r^2/4}{}_1F_1\left(-\dfrac{\nu}{2 \lambda},M+1; - \f{\lam r^2}{2}\right)\ea\right),~~\lambda < 0,~~M = 0, 1, 2 \cdots .
\eeq
We would like to mention that the exact zero energy solutions are a result of chirality. The chirality operator is given by \cite{chiral} $\s_z$ and it is easy to see that the solutions (\ref{zero1}), (\ref{zero3}) have chirality $+1$ while the solutions (\ref{zero2}), (\ref{zero4}) have chirality $-1$.
\section{Conclusion}\label{con}
In this paper we have considered generalized harmonic confinement of quasiparticles in graphene. The magnetic fields used for confining the quasiparticles are finite everywhere. More specifically, we have considered magnetic fields which are of a non polynomial type and is expressed in terms of confluent hypergeometric series or Laguerre polynomials. The conditions required to produce the bound states have also been analyzed in detail. In this context we would like to point out that for generation of magnetic fields without any singularity parameters have to chosen in such a way that the confluent hypergeometric series/associated Laguerre polynomials remain free of zeros. Following Refs. \onlinecite{negative,sansone,ahmed} one may find necessary conditions for absence of zeros of the above functions. Here it has been shown in the examples considered that by judicious choices of the parameters it is indeed possible to create finite magnetic fields (and hence confinement). We would also like to point out that for the types of magnetic field considered here confinement can be achieved for either for $m\geq 0$ or for $m<0$ (equivalently $M\leq -1$ and $M>-1$) but for a particular magnetic field all values of angular momentum are not allowed. It is also of interest to note that in some cases the pseudospinors could also be expressed in terms of the recently discovered exceptional orthogonal polynomials. 
This also establishes a relation between the magnetic field profiles and exceptional orthogonal polynomials. It may be mentioned that although we have not discussed any specific experiment, nevertheless we feel that finite magnetic fields considered in this paper may be produced following the suggestions in \cite{downing1,downing2}.

In this paper we have considered the massless $(2+1)$ dimensional Dirac equation. It may be mentioned that for gapped graphene or other similar Dirac materials the electrons or quasiparticles are not massless and a mass term of the form $\Delta \s_z$ has to be added in Eq.(\ref{eqn:Hamilton}). In such a case the procedure to obtain the spectrum and the corresponding solutions would remain largely the same. Finally it may be noted that here we have searched for magnetic fields for which the effective potentials are generalization of two dimensional harmonic oscillators. We believe it would be of interest to search for other types of magnetic fields which may produce effective potentials of non harmonic types.
\begin{acknowledgments}
The authors like to thank the referee for useful suggestions. Two of the authors (Dai-Nam Le and Van-Hoang Le) are supported by the Vietnam National Foundation for Science and Technology Development (NAFOSTED) under Grant No. 103.01-2017.371.
\end{acknowledgments}

\end{document}